\documentclass[]{aa}
\usepackage{psfig,latexsym,longtable,lscape}

\def\degmark{^\circ}
\def \rsun {\ifmmode$R$_{\odot}\else R$_{\odot}$\fi}
\def \nh {N${\rm _H}$}
\def \hcm {\hbox {\ifmmode $ H atoms cm$^{-2}\else H atoms cm$^{-2}$\fi}}
\def \src {1E~0035.4-7230}
\def\approxgt{\mathrel{\hbox{\rlap{\lower.55ex \hbox {$\sim$}}
        \kern-.3em \raise.4ex \hbox{$>$}}}}
\def\approxlt{\mathrel{\hbox{\rlap{\lower.55ex \hbox {$\sim$}}
        \kern-.3em \raise.4ex \hbox{$<$}}}}

\newcommand {\einstein} {{\it Einstein}}

\newcommand {\sax} {{\it BeppoSAX}}

\newcommand {\degree} {$^{\circ}$}

\def \nh {N${\rm _H}$}

\begin{document}

\thesaurus{ (02.01.2; 08.02.1; 08.09.2; 08.23.1; 13.25.5)}

\title{A BeppoSAX observation of the supersoft source 1E~0035.4-7230}

\author{P. Kahabka\inst{1}, A.N. Parmar\inst{2}, \and H.W. Hartmann\inst{3}}

\institute{Astronomical Institute and Center for High Energy Astrophysics,
           University of Amsterdam, Kruislaan 403, 1098 SJ Amsterdam, The
           Netherlands 
\and
           Astrophysics Division, Space Science Department of ESA, ESTEC,
           P.O. Box 299, 2200 AG Noordwijk, The Netherlands 
\and
           SRON Laboratory for Space Research, Sorbonnelaan 2, 
           3584 CA Utrecht, The Netherlands
         }

\date{Received 17 February 1999 / accepted 30 March 1999}
\offprints{ptk@astro.uva.nl}

\maketitle\markboth{P. Kahabka et al.: BeppoSAX observation of 1E~0035.4-7230}
                   {P. Kahabka et al.: BeppoSAX observation of 1E~0035.4-7230}

\begin{abstract}
Results from a 37~ks BeppoSAX Low-Energy Concentrator Spectrometer (LECS) 
observation of the supersoft source SMC~13 (=1E~0035.4-7230) in the Small 
Magellanic Cloud are reported. This source has probably the softest spectrum 
observed so far with BeppoSAX, with no detected counts $\approxgt$0.5~keV. 
The BeppoSAX spectrum is fitted either with a blackbody spectrum with an
effective temperature kT~=~26--58~eV, an LTE white dwarf atmosphere spectrum
with kT~=~35--50~eV, or a non-LTE white dwarf atmosphere spectrum with 
kT~=~25--32~eV. The bolometric luminosity is not very well constrained, it
is $\rm < 8\ 10^{37}\ erg\ s^{-1}$ and $\rm < 3\ 10^{37}\ erg\ s^{-1}$ for
the LTE and the non-LTE spectrum (90\% confidence). 
 
We also applied a spectral fit to combined spectra obtained with BeppoSAX 
LECS and with ROSAT PSPC. We find that a blackbody spectrum with an effective 
temperature kT=(39--47)~eV and a bolometric luminosity of $\rm (0.3-5)\ 
10^{37}\ erg\ s^{-1}$ fits the data. The data are also fitted with a blackbody
with a kT of (50--81)~eV, an average C-edge at (0.38--0.47)~keV with an
optical depth $\rm \tau>1.1$, and a bolometric luminosity of $\rm (3-8)\ 
10^{36}\ erg\ s^{-1}$ (90\% confidence). We also applied LTE and non-LTE 
white dwarf atmosphere spectra. The kT derived for the LTE spectrum is 
(45--49)~eV, the bolometric luminosity is $\rm (3-7)\ 10^{36}\ erg\ s^{-1}$,
The kT derived for the non-LTE spectrum is (27--29)~eV, the bolometric 
luminosity is $\rm (1.1-1.2)\ 10^{37}\ erg\ s^{-1}$. We can exclude any 
spectrally hard component with a luminosity of more than $\rm 2\ 10^{35}\ 
erg\ s^{-1}$ (for a bremmstrahlung with a temperature of 0.5~keV) at a 
distance of 60~kpc. The LTE temperature is therefore in the range 
$\rm 5.5\pm0.2\ 10^5$~K and the non-LTE temperature in the range 
$\rm 3.25\pm0.16\ 10^5$~K.

Assuming the source is on the stability line for atmospheric nuclear burning, 
we constrain the white dwarf mass from the LTE and the non-LTE fit to 
$\sim$1.1~${\rm M_{\odot}}$ and $\sim$0.9~${\rm M_{\odot}}$ respectively.
However, the temperature and luminosity derived with the non-LTE model for 
1E~0035.4-7230 is consistent with a lower mass (${\rm M_{\rm WD} \sim 0.6-0.7 
M_{\odot}}$) white dwarf as predicted by Sion \& Starrfield (1994). At the 
moment, neither of these two alternatives for the white dwarf mass can be 
excluded.
\end{abstract}


\section{Introduction}
The \einstein\ observatory performed a survey of the Small Magellanic Cloud 
(SMC) in which two sources with unusually soft spectra, 1E~0056.8-7154 and 
1E~0035.4-7230, were detected (Seward \& Mitchell 1981). 1E~0035.4-7230
has been identified in the optical with a variable blue star with a strong UV 
excess and a 4.1~hour orbital period (Orio et al. 1994). The supersoft nature 
of the source was confirmed by subsequent ROSAT observations (Kahabka et al. 
1994; Kahabka 1996). Supersoft sources are most probably accreting white 
dwarfs (WDs) which are burning hydrogen steadily (van den Heuvel et al. 1992). 
An orbital period of 0.1719~days was found in optical and X-ray data (Crampton
et al. 1997; see van Teeseling et al. 1998 for a recent discussion). The 
optical spectrum shows pronounced emission at He{\sc ii}~$\lambda$4686 and 
near H$\alpha$. The H$\beta$ line is seen in absorption (Cowley et al. 1998). 
The mass function derived from the He{\sc ii}~$\lambda$4686 line radial 
velocities indicates a donor star mass of $\sim$${\rm 0.4\ M_{\odot}}$ and a 
WD mass of $\sim$${\rm0.5-1.4\ M_{\odot}}$. A scenario to account for the 
evolutionary state of the source has been proposed by Kahabka \& Ergma (1997) 
in which the system is a cataclysmic variable (CV) with a low-mass WD 
accreting below the steady-state burning rate. The recurrent flashes in this 
system are assumed to be mild and leave a fraction of the envelope as a 
nuclear reservoir. The WD is heated to a high temperature during its evolution 
(Sion \& Starrfield 1994). The Sion \& Starrfield models of a ${\rm 0.6\ 
M_{\odot}}$ and a ${\rm 0.7\ M_{\odot}}$ WD predict an asymptotic temperature 
of $\sim$$\rm 3.25\ 10^5~$K and a luminosity of $\sim$$\rm 
10^{37}$~erg~s$^{-1}$. X-ray spectroscopy allows the WD mass and temperature 
to be estimated, so allowing the evolutionary status of the system to be 
probed. If 1E~0035.4-7230 harbors a massive WD (${\rm M_{\rm WD} > 0.7-0.8\ 
M_{\odot}}$) then the Sion \& Starrfield model may no longer be applicable as 
the envelope is probably expelled during recurrent nova outbursts. This work 
is part of a program to systematically observe the brighter supersoft sources 
with the BeppoSAX satellite (Parmar et al. 1997a, 1998; Hartmann et al. 1999).

\section{Observation}
The Low-Energy Concentrator Spectrometer (LECS) onboard \sax\ is an imaging 
gas scintillation proportional counter sensitive in the energy range 
0.1--10.0~keV with a circular field of view of 37$'$ diameter (Parmar et al. 
1997b). Its energy resolution is 
a factor $\sim$2 better than that of the ROSAT Position Sensitive 
Proportional Counter (PSPC).
\src\ was observed between 1998 January 4 10:27 and January 6 01:21~UTC. 
Good data were selected from intervals when the minimum elevation angle 
above the Earth's limb was $>$5$\degmark$ and when the high voltage levels 
were nominal using the SAXDAS data analysis package. Since the LECS was only 
operated during satellite night-time, this gave a total on-source exposure 
of 37~ks. 

Examination of the LECS image shows a source at a position consistent with 
that of 1E~0035.4-7230. A spectrum was extracted centered on the source 
centroid using a radius of 8$'$. This radius was chosen to include 95\% of 
the 0.28~keV photons. The spectrum was rebinned to have $>$20 counts in each 
bin to allow the use of the $\rm \chi^2$ statistic. The LECS response matrix 
from the 1997 September release was used in the spectral analysis. The 
background spectrum was extracted from the image itself using a semi-annulus 
centered on 1E~0035.4-7230 with inner and outer radii of 14$'$ and 19$'$, 
respectively. A correction for telescope vignetting was applied (see Parmar et
al. 1999). The \src\ count rate above background is 0.008~s$\rm ^{-1}$. 
Examination of the extracted spectrum shows that the source is only detected 
in a narrow energy range and only the 21 rebinned channels corresponding to 
energies between 0.13 and 1.7~keV were used for spectral fitting. 

\section{BeppoSAX spectral fit}
Initially only BeppoSAX data was used to constrain the parameters derived from
different spectral models. The fits indicate the presence of a C edge at 
$\sim$0.5~keV, most likely due to C~{\sc vi} and this provides strong support 
for a hot and nuclear burning WD atmosphere model. 

\begin{figure}[htbp]
  \centering{ 
  \vbox{\psfig{figure=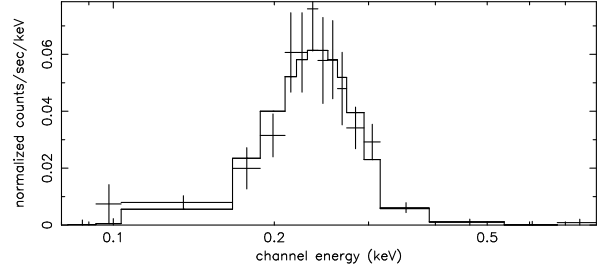,width=8.0cm,%
  bbllx=1.0cm,bblly=0.5cm,bburx=22.0cm,bbury=10.1cm,clip=}}\par
            }
  \caption[]{BeppoSAX count spectrum of the 37~ks observation of 
             1E~0035.4-7230 together with the best-fit non-LTE spectral
             fit}
\end{figure}

\begin{figure}[htbp] 
  \centering{ 
  \vbox{\psfig{figure=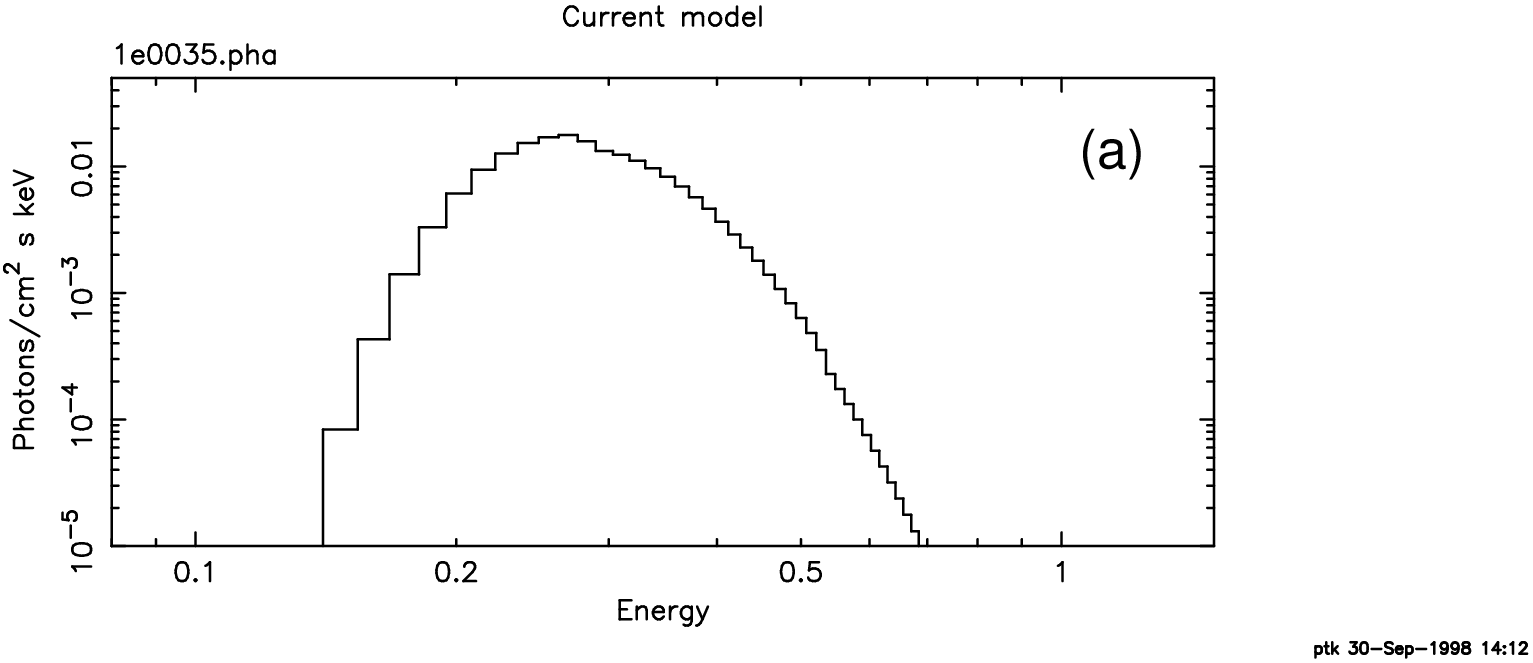,width=7.0cm,%
  bbllx=1.0cm,bblly=0.5cm,bburx=22.0cm,bbury=10.0cm,clip=}}\par
    \vbox{\psfig{figure=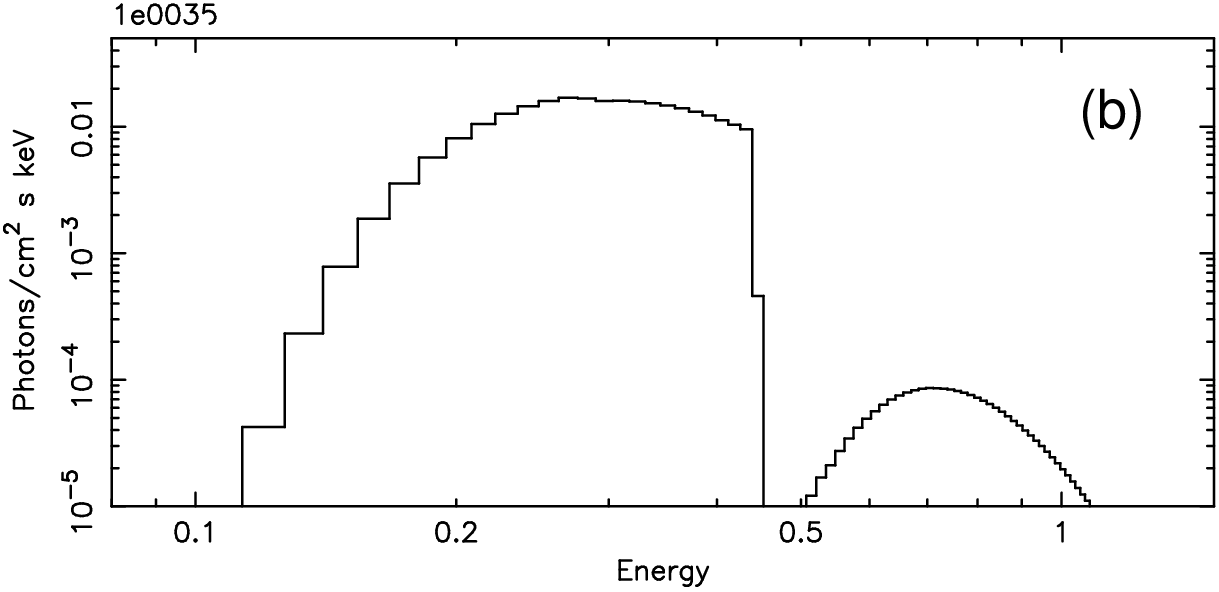,width=7.0cm,%
  bbllx=1.0cm,bblly=0.5cm,bburx=22.0cm,bbury=10.0cm,clip=}}\par
  \vbox{\psfig{figure=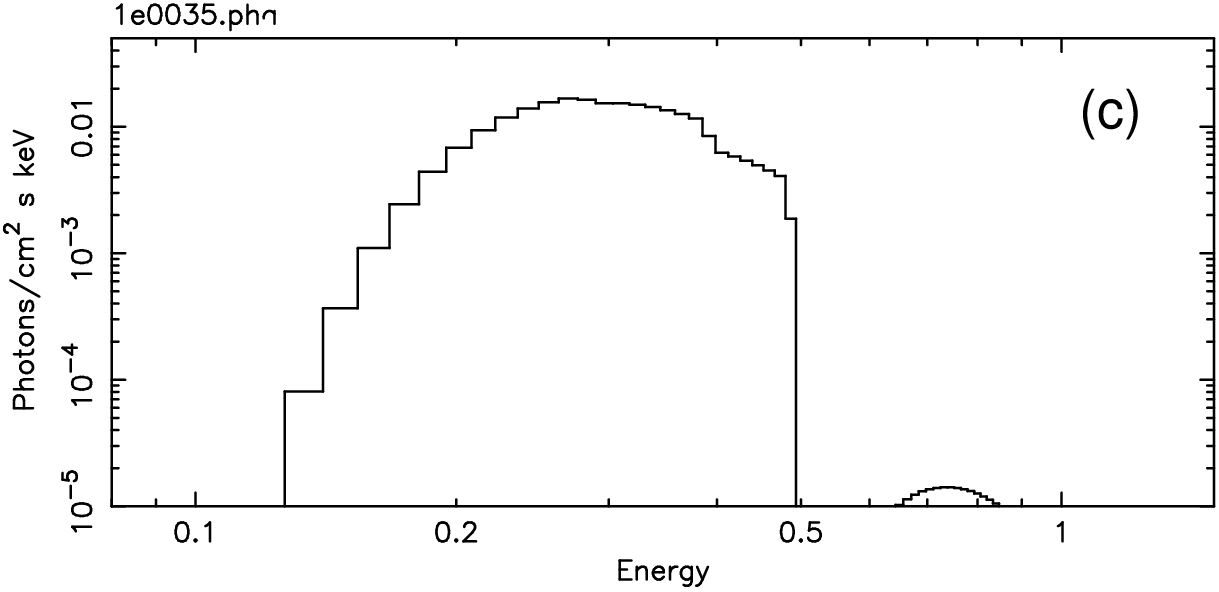,width=7.0cm,%
  bbllx=1.0cm,bblly=0.5cm,bburx=22.0cm,bbury=10.0cm,clip=}}\par
  \vbox{\psfig{figure=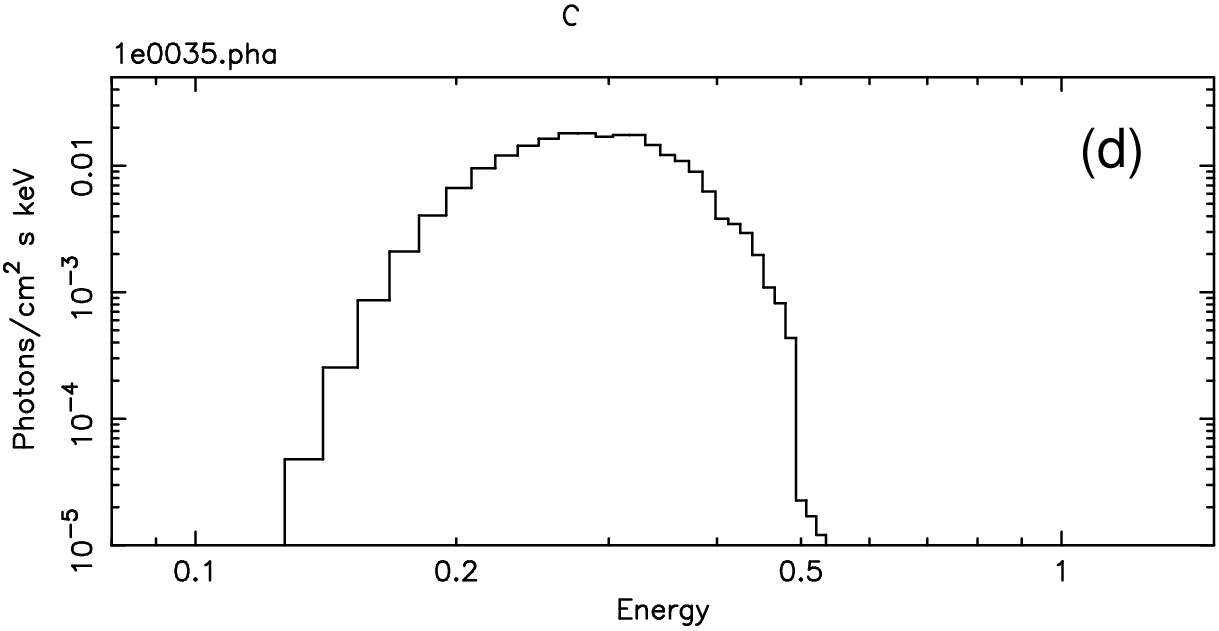,width=7.0cm,%
  bbllx=1.0cm,bblly=0.5cm,bburx=22.0cm,bbury=10.0cm,clip=}}\par
  \vbox{\psfig{figure=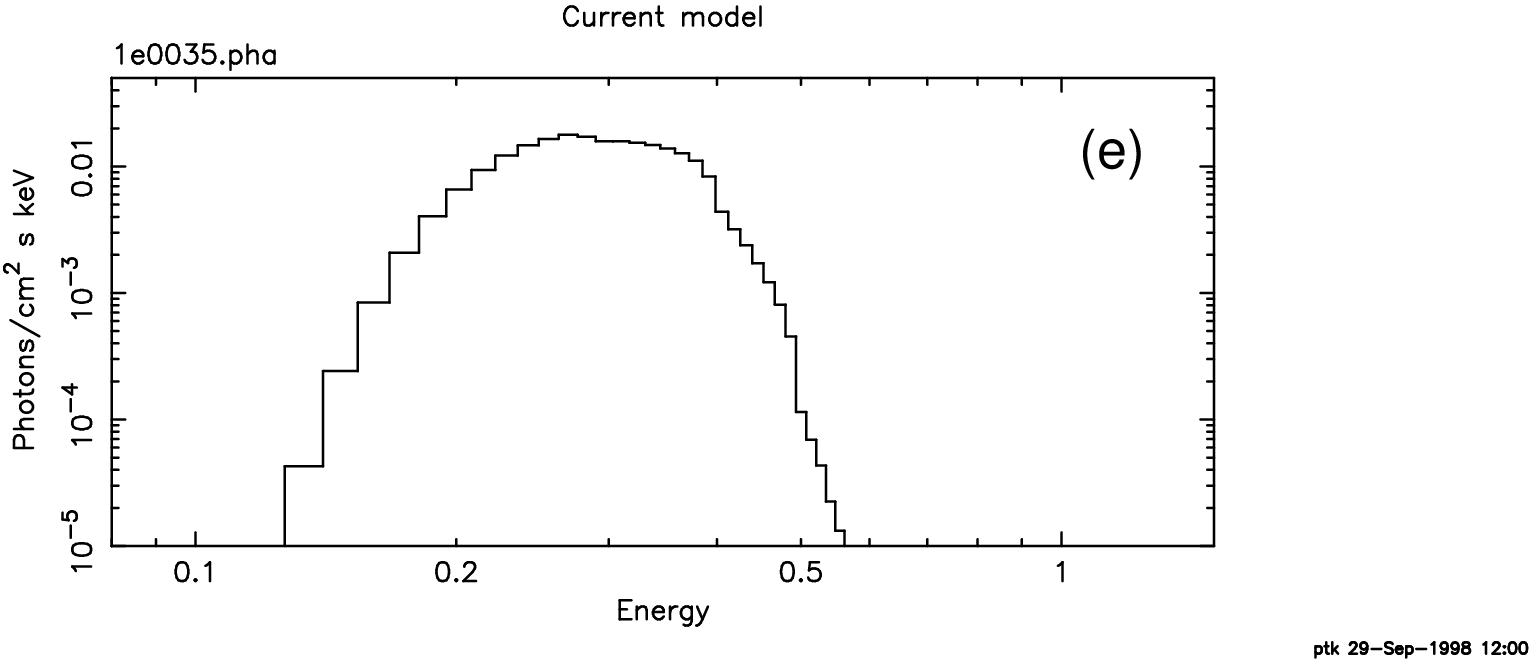,width=7.0cm,%
  bbllx=1.0cm,bblly=0.5cm,bburx=22.0cm,bbury=10.0cm,clip=}}\par
            }
  \caption[]{(a) Best-fit blackbody model. The blackbody kT is 39~eV and 
             the column ${\rm 4.7\ 10^{20}}$ ${\rm H-atoms~cm^{-2}}$.(b) 
             Best-fit 
             blackbody model with an edge at 0.44~keV. The optical depth is 
             larger than 1.8 (68\% confidence). The blackbody kT is 79~eV and 
             the column ${\rm 2.2\ 10^{20}}$ ${\rm H-atoms~cm^{-2}}$. (c) 
             Best-fit 
             blackbody model with a C~{\sc v} edge at 0.391~keV and a C~{\sc 
             vi} edge at 0.487~keV. The optical depths are 0.4 and $>$1.3, 
             respectively. The blackbody kT is 63~eV and \nh\ ${\rm 3.1\ 
             10^{20}}$ ${\rm H-atoms~cm^{-2}}$. (d) Best-fit LTE WD atmosphere
             model.
             The kT is $\rm 5.5\ 10^5$~K and \nh\ ${\rm 3.3\ 10^{20}}$
             ${\rm H-atoms~cm^{-2}}$. (e) Best-fit non-LTE WD atmosphere model.
             The 
             kT is $3.9\ 10^5$~K and \nh\ ${\rm 3.5\ 10^{20}}$ 
             ${\rm H-atoms~cm^{-2}}$}
\end{figure}

\subsection{Blackbody spectral fits}

The BeppoSAX data were first fit with a blackbody spectral model. The fit 
is acceptable with a $\chi^2$ of 8.5 for 18 degrees of freedom (dof). The 
blackbody temperature, kT, is (20--52)~eV and the column, \nh, ${\rm (2.2-11)\
10^{20}}$ ${\rm H-atoms~cm^{-2}}$. The best-fit luminosity (60~kpc) is ${\rm 
10^{37}\ erg\ s^{-1}}$, but it is not well constrained (cf. Tab.1).

Next edges were included in the spectral model.

\begin{equation}{\rm
  Bbody \times f(E)}
\end{equation} 

where 

\begin{equation}{\rm
f(E) = \left\{ \begin{array}{c@{\quad:\quad}l}
 1.                                             & E<E_{\rm edge} \\
 exp(-\tau \times ({\frac{E}{E_{\rm edge}}})^3) & E\ge E_{\rm edge}
              \end{array} \right.
             }
\end{equation} 

Edges are expected to be present in the atmospheres of hot WDs. Calculations 
of synthetic spectra have shown that such edges are the dominant spectral 
features in this energy range. These are likely to be Lyman edges, i.e. 
ionizations from the ground state of C~{\sc v} and C~{\sc vi} (at energies of 
0.391 and 0.487~keV). A single edge with variable energy was first included. 
As the fit did not allow to constrain the energy of the edge we fixed the edge
energy to 0.44~keV, the mean expected energy of the C~{\sc v} and the 
C~{\sc vi} edge of 0.439~keV. The optical depth, $\tau>1.8$, at the 68\% 
confidence level. At the 90\% confidence level $\tau$ cannot be constrained. 
The $\chi^2$ is 7.1 for 17 dof. The best-fit blackbody kT is 79~eV. The \nh\ 
of ${\rm (0.8-8.8)\ 10^{20}}$ ${\rm H-atoms~cm^{-2}}$ is in agreement with the 
foreground column of ${\rm 3.2\ 10^{20}}$ ${\rm H-atoms~cm^{-2}}$ (derived 
from the 
21-cm map of Stanimirovic et al. 1999). The luminosity is ${\rm (0.7-8.4)\ 
10^{36}\ erg\ s^{-1}}$. We applied the F-test statistics and find a 
probability of $\sim$70\% that a fit including a C-edge is more likely than 
a fit without an edge (leaving the blackbody temperature in the fit as a free 
parameter). 

Next edges fixed at the energies of the C~{\sc v} and C~{\sc vi} edges were 
included in the spectral model. The best-fit optical depths are
${\rm \tau_{\rm C~V}\le 1.8}$ and ${\rm \tau_{\rm C~VI}\ge 1.3}$
at the 68\% confidence level (${\rm \tau_{\rm C~V}\le 3.0}$ at the 90\%
confidence level). The best-fit blackbody kT is 63~eV. The \nh\ is 
${\rm (2.0-6.1)\ 10^{20}}$ ${\rm H-atoms~cm^{-2}}$, and is in agreement with 
the galactic foreground column. The luminosity is ${\rm (1.0-9.6)\ 10^{36}\ 
erg\ s^{-1}}$.

\subsection{WD atmosphere spectral fits}

A Blackbody spectrum including absorption edges of C~{\sc v} and C~{\sc vi}
can only be an approximation to the true spectral distribution. LTE and
non-LTE WD atmosphere spectra may be the better approximation (Heise et al. 
1994; Hartmann \& Heise 1997). We therefore applied LTE and non-LTE WD 
atmosphere spectra to the BeppoSAX spectrum of 1E~0035.4-7230 assuming 
reduced abundances for C, N, O and Ne, also in Section 4.2.

The LTE fit is acceptable with a $\chi^2$ of 8.0 for 17 dof. The effective kT 
is $(4.1-5.9)\ 10^5$~K, the \nh\ is ${\rm (1.6-6.5)\ 10^{20}}$ 
${\rm H-atoms~cm^{-2}}$ and the best-fit luminosity ${\rm 3.4\ 10^{36}\ erg\ 
s^{-1}}$, the luminosity is not well constrained. The non-LTE fit is also 
acceptable with a $\chi^2$ of 8.0 for 17 dof. The best-fit effective kT of 
$(3.3 \pm _{0.3} ^{0.5})\ 10^5$~K is significantly lower than that derived 
from the LTE fit. The \nh\ is ${\rm (2.3-6.6)\ 10^{20}}$ ${\rm H-atoms\ 
cm^{-2}}$. The luminosity is in the range  ${\rm (0.5-3.8)\ 10^{37}\ erg\ 
s^{-1}}$.

\begin{table*}
  \caption[]{Results of spectral fitting of the BeppoSAX~LECS spectrum 
  (0.1--1.7~keV) and combined BeppoSAX~LECS and ROSAT~PSPC spectrum of 
  1E~0035.4-7230. The (0.1--2.4)~keV luminosity L$_{36}$ ($\rm 10^{36}\ 
  erg\ s^{-1}$) and for the combined ROSAT~PSPC and BeppoSAX~LECS spectral 
  fit also the bolometric luminosity L$_{36}$ ($\rm 10^{36}\ erg\ s^{-1}$) 
  is given. 90\% confidence parameter ranges are given ($^{a}$ 68\% confidence
  parameter range is given).}
  \begin{flushleft}
  \begin{tabular}{lccccccccc}
  \hline
  \noalign{\smallskip}
  \multicolumn{9}{c}{BeppoSAX LECS} \\
  \noalign{\smallskip}
  Model                 &\nh&kT&${\rm E_{\rm C~{V}}}$&${\rm E_{\rm C~{VI}}}$
&${\rm \tau_{\rm C~V}}$&${\rm \tau_{\rm C~VI}}$ & L$_{36}$ && $\chi$/dof        \\
                        &             &      & \multicolumn{4}{c}{Edge} 
&        (0.1-2.4~keV)&                                    \\
                        & ($10^{20}\ cm^{-2}$) &(eV; &\multicolumn{2}{c}
{(keV)}
& & & &       \\
                        &               &$10^5$~K) &        &                
& & &                          &       \\
  \noalign{\smallskip}
  \hline
  \noalign{\smallskip}
 bbody     &
$4.7\pm_{2.5}^{6.8}$&$39\pm_{19}^{13}$&               &                     &              &                &$10\pm_{9.3}^{>10}$     &&0.47\\
          &         &$4.5\pm_{2.2}^{1.5}$&                                       &                          &                    &&    \\
 \noalign{\smallskip}
 bbody+edge&
$2.2\pm_{1.4}^{6.6}$&$79\pm_{55}^{19}$&\multicolumn{2}{c}{0.44}&\multicolumn{2}{c}{$>1.8^{a}$}&$2.2\pm_{1.5}^{6.2}$&&0.42\\
          &         &$9.2\pm_{6.3}^{2.2}$&                                       &                          &                    &&    \\
 \noalign{\smallskip}
 bbody+C{\sc v}+C{\sc vi} edge&
$3.1\pm_{2.1}^{3.0}$&$63\pm_{23}^{33}$&0.391&0.487&$0.4\pm_{0.4}^{2.6}$  &$>1.3^{a}$                    &$2.8\pm_{1.8}^{6.8}$      &&0.57\\
         &          &$7.3\pm_{2.2}^{2.9}$&                                       &                          &                    &&    \\
 \noalign{\smallskip}
 LTE WD atm.    &
$3.3\pm_{1.7}^{3.2}$&$47\pm_{12}^{3}$&  &            &                            &                          &$3.4\pm_{3.1}^{82}$&&0.47\\
         &          &$5.5\pm_{1.4}^{0.4}$&                                       &                          &                    &&    \\
 \noalign{\smallskip}
 Non-LTE WD atm.&
   $4.0\pm_{1.7}^{2.6}$&$28\pm_{3}^{4}$& &     &                        &                          &$8.6\pm_{3}^{29}$ &&0.47\\
         &             &$3.3\pm_{0.3}^{0.5}$& &    &                        &                      &  &&   \\
 \noalign{\smallskip}
  \noalign{\smallskip}
  \hline
  \noalign{\smallskip}
  \multicolumn{9}{c}{BeppoSAX LECS + ROSAT PSPC} \\
  \noalign{\smallskip}
 Model          & 
   \nh           & kT & ${\rm E_{C~{V}}}$ & ${\rm E_{C~{VI}}}$ & ${\rm \tau_{\rm C~V}}$ & ${\rm \tau_{\rm C~VI}}$ & L$^{c}_{36}$ &scaling& $\chi$/dof\\
                &  &  & \multicolumn{4}{c}{Edge}                                                        &(0.1-2.4~keV)&factor$^{b}$&\\
   & ($10^{20}\ cm^{-2}$) &(eV; &\multicolumn{2}{c}{(keV)}& & & (bolometric)                           &&           \\
                        &               &$10^5$~K) &        &                                 & & &                      &&           \\
  \noalign{\smallskip}
  \hline
  \noalign{\smallskip}
 bbody             &
$4.8\pm_{0.8}^{1.1}$ &$43\pm_{4}^{4}$&      &     &                        &                          &$9.3\pm_{4.2}^{11}$&$1.68\pm_{0.19}^{0.23}$&0.68     \\
                  &  &$5.0\pm_{0.5}^{0.5}$ &&     &                        &                          &$12\pm_{9}^{42}$&\\ 
 \noalign{\smallskip}
 bbody+edge      &
$3.8\pm_{0.8}^{1.1}$&$63\pm_{15}^{13}$&\multicolumn{2}{c}{$0.44\pm_{0.08}^{0.03}$}&\multicolumn{2}{c}{$5.4{\pm_{2.7}^{3.6}}^{a}$}&$4.2\pm_{1.5}^{3.8}$&$1.67\pm_{0.17}^{0.23}$&0.62\\
                  &  &$7.3\pm_{1.7}^{1.5}$ &&     & & &$5.2\pm_{2.0}^{0.9}$   &                          &                               \\ 
 \noalign{\smallskip}
 bbody+C{\sc v}+C{\sc vi} edge &
   $3.9\pm_{0.8}^{1.2}$ &$65\pm_{15}^{16}$      &0.391&0.487                   &$1.6{\pm_{0.7}^{1.0}}^{a}$    &$3.9{\pm_{2.8}^{3.6}}^{a}$&$4.2\pm_{1.4}^{3.3}$&$1.68\pm_{0.12}^{0.14}$&0.63\\
                  &  &$7.5\pm_{1.7}^{1.5}$ &&     &                        &             &$5.4{\pm_{3.5}^{3.6}}$ & &            \\ 
 \noalign{\smallskip}
 LTE WD atm.$^{d}$    &
   $3.6\pm_{0.4}^{0.4}$    &$47\pm_{1}^{1}$ &     &                        &                          & &$2.0\pm_{0.8}^{1.4}$&$1.68\pm_{0.1}^{0.2}$&0.80\\
                  &  &$5.5\pm_{0.1}^{0.1}$  &     &                        &                          & &   $4.8\pm_{0.8}^{1.1}$& \\ 
 \noalign{\smallskip}
 Non-LTE WD atm.$^{d}$&
   $4.1\pm_{0.3}^{0.4}$ &$28\pm_{0.8}^{0.6}$&     &                        &                          & &$9.8\pm_{1.4}^{4.0}$&$1.30\pm_{0.14}^{0.15}$&0.77\\
                  &  &$3.25\pm_{0.09}^{0.07}$&    &                        &                          & &$10.9\pm_{0.3}^{0.3}$&\\ 
 \noalign{\smallskip}
  \noalign{\smallskip}
  \hline
  \end{tabular}
  \end{flushleft}
$^{b}$ the flux scaling factor is ($\rm flux_{ROSAT PSPC}$ / ($\rm flux_{BeppoSAX}$), $^{c}$ the luminosity is given for the BeppoSAX spectrum. $^{d}$ 68\% confidence errors are given. In the text estimates for the 90\% errors are given as twice the 68\%
errors. 
\end{table*} 

\begin{figure}[htbp]
  \centering{ 
  \vbox{\psfig{figure=8569_f3a.ps,width=7.0cm,%
  bbllx=1.0cm,bblly=0.5cm,bburx=22.0cm,bbury=10.0cm,clip=}}\par
    \vbox{\psfig{figure=8569_f3b.ps,width=7.0cm,%
  bbllx=1.0cm,bblly=0.5cm,bburx=22.0cm,bbury=10.0cm,clip=}}\par
  \vbox{\psfig{figure=8569_f3c.ps,width=7.0cm,%
  bbllx=1.0cm,bblly=0.5cm,bburx=22.0cm,bbury=10.0cm,clip=}}\par
  \vbox{\psfig{figure=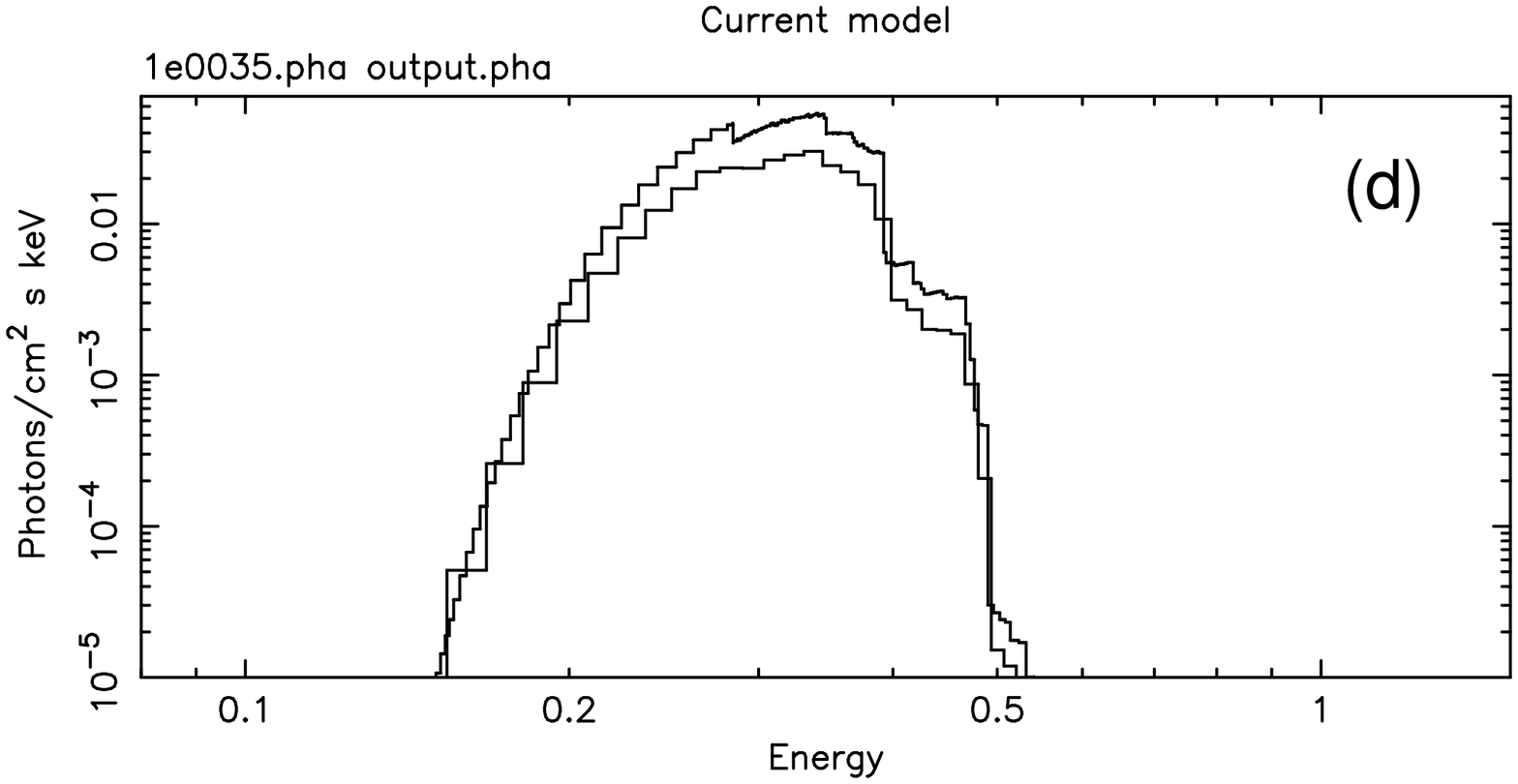,width=7.0cm,%
  bbllx=1.0cm,bblly=0.5cm,bburx=22.0cm,bbury=10.0cm,clip=}}\par
  \vbox{\psfig{figure=8569_f3e.ps,width=7.0cm,%
  bbllx=1.0cm,bblly=0.5cm,bburx=22.0cm,bbury=10.0cm,clip=}}\par
            }
  \caption[]{(a) Best-fit blackbody model. The blackbody kT is 43~eV and
             \nh\ is ${\rm 4.9\ 10^{20}}$ ${\rm H-atoms~cm^{-2}}$. (b) 
             Best-fit 
             blackbody model with an edge at 0.44~keV. The optical depth 
             is 5.4. The blackbody kT is 63~eV and \nh\ is ${\rm 3.8\ 
             10^{20}}$ ${\rm H-atoms~cm^{-2}}$. (c) Best-fit blackbody model 
             with a C~{\sc v} edge at 0.391~keV and a C~{\sc vi} edge at 
             0.487~keV. The optical depths are 1.6 and 3.9, respectively. The 
             blackbody kT is 65~eV and \nh\ is ${\rm 3.9\ 10^{20}}$ 
             ${\rm H-atoms~cm^{-2}}$. (d) Best-fit LTE WD atmosphere model. 
             The kT is $5.5\ 10^5$~K, and the \nh\ ${\rm 3.6\ 10^{20}}$ 
             ${\rm H-atoms~cm^{-2}}$. (e) Best-fit non-LTE WD atmosphere 
             model. The kT is $3.25\ 10^5$~K, and the \nh\ ${\rm 4.1\ 
             10^{20}}$ ${\rm H-atoms~cm^{-2}}$.}
\end{figure}

\section{Combined BeppoSAX and ROSAT PSPC spectral fits}

In a second step we used the BeppoSAX spectrum together with a ROSAT spectrum 
obtained from a 13~ks exposure in 1993 April in a combined (multi-instrument) 
spectral fit to better constrain the spectral parameters. Fits to the ROSAT 
data alone are presented in Kahabka et al. (1994). Again the fits indicate the 
presence of a C edge at $\sim$0.5~keV, most likely due to C~{\sc vi} and are 
in strong support for a hot and nuclear burning WD nature of the source.

\subsection{Blackbody spectral fits}

The data were first fit with a blackbody spectral model. The fit is acceptable 
with a $\chi^2$ of 40 for 59 dof. The blackbody kT is (39--47)~eV, the \nh\ of
${\rm (4.1-6.0)\ 10^{20}}$ ${\rm H-atoms~cm^{-2}}$ is greater than the 
foreground column of ${\rm 3.2\ 10^{20}}$ ${\rm H-atoms~cm^{-2}}$. The 
bolometric 
luminosity is ${\rm (1-5)\ 10^{37}\ erg\ s^{-1}}$. There is no indication of 
the existence of neutral circumstellar matter (e.g. a nebula) in this system. 
Any low density 
matter in the binary system (mass-loss due to winds) found from the modeling 
of the line profiles of the Balmer lines has been shown to be highly ionized 
(cf. Asai et al. 1998 for the supersoft source CAL~87 located in the LMC). 

Next a single edge with variable energy and depth was included in the spectral 
model. The fit is acceptable with a $\chi^2$ of 35 for 56 dof. The best-fit 
indicates an edge energy in the range (0.36--0.47)~keV with a best-fit value 
of 0.44~keV. This value is consistent with the mean expected energy of the 
C~{\sc v} and the C~{\sc vi} edge (at 0.391~keV and 0.487~keV, respectively) 
which is 0.439~keV. The derived energy range indicates an ionization state to 
be somewhere between C~{\sc v} and C~{\sc vi}. The optical depth is ${\rm 
\tau_{\rm C}}>1.1$ at the 90\% confidence level. In this model the blackbody 
temperature is found to be 48--76~eV. The \nh\ of ${\rm (3.0-5.0)\ 10^{20}}$
${\rm H-atoms~cm^{-2}}$ is in agreement with the foreground column. We applied
the 
F-test statistics and find a probability of $\sim$70\% that a fit including a 
C-edge is more likely than a fit without an edge (leaving the blackbody 
temperature in the fit as a free parameter). As before, the edge energies were
next fixed at the values appropriate to C~{\sc v} and C~{\sc vi} edges in the 
spectral fit. The fit is acceptable with a $\chi^2$ of 35 for 56 dof. The 
best-fit optical depths are ${\rm \tau_{\rm C~V}}$= 0.9--2.6 and 
${\rm \tau_{\rm C~VI}}$=1.1--7.5 at 68\% confidence, ${\rm \tau_{\rm 
C~V}}<2.9$ at 90\% confidence. The blackbody kT is in the range 
50--81~eV. Using the Saha equation we may constrain the electron density 
(cf. Sect.~5.1). The \nh\ of ${\rm (3.1-5.1)\ 10^{20}}$ ${\rm H-atoms\ 
cm^{-2}}$ is in agreement with the galactic foreground column.

\subsection{WD atmosphere spectral fits}

The LTE fit assuming reduced abundances is acceptable with a $\chi^2$ of 47 
for 59 dof. The effective kT is $(5.5\pm 0.2)\ 10^5$~K, the \nh\ is ${\rm 
(3.6 \pm 0.8) 10^{20}}$  ${\rm H-atoms\ cm^{-2}}$, the bolometric luminosity 
is ${\rm (3-7)\ 10^{36}\ erg\ s^{-1}}$. With non-LTE WD atmosphere spectrum 
and assuming reduced abundances (cf. section~3.2) significantly lower kT of 
$(3.25 \pm 0.16)\ 10^5$~K is found. The derived \nh\ is ${\rm (4.1 \pm0.8)\ 
10^{20}}$ ${\rm H-atoms~cm^{-2}}$. The fit is acceptable with a $\chi^2$ of 
46 for 58 dof. The bolometric luminosity is $\rm (1.0-1.2)\ 10^{37}\ erg\ 
s^{-1}$.

\section{Discussion}

We find strong evidence in the BeppoSAX and ROSAT spectra for the presence of 
an edge at $\sim$0.44~keV in the spectrum of 1E~0035.4-7230. This edge 
originates from a combination of C~{\sc v} and C~{\sc vi} with a large optical 
depth $\tau > 1.1$ (90\% confidence).

The temperatures and absorbing
columns derived from the blackbody fits should be treated with caution and 
more reliable estimates are probably obtained with the LTE or non-LTE WD 
atmosphere model fits. The resulting chi-squared values do not allow to decide
between both models. Therefore, the temperature is constrained to the range 
$(3-6)\ 10^5$~K and the ionization state of C is likely between C~{\sc v} and 
C~{\sc vi}. The discovery of deep C edges in the spectrum of 1E~0035.4-7230 
has implications as it is assumed that the metallicity of the SMC material is 
heavily reduced (by a factor 3--20, Pagel 1993). Reduced CNO abundances result
in less violent outbursts. This can sustain nuclear burning for long periods
of time without the requirement of additional supply due to accretion e.g. 
by X-ray induced heating of the donor star (van Teeseling \& King 1998; 
King \& van Teeseling 1998). It is unclear what enhances the C abundance in 
the envelope of the WD. If the WD goes through recurrent weak outbursts then 
the CO core may be affected (i.e. eroded) during the outburst. Eroded core 
material (i.e. C and O) may be mixed into the envelope. But as no N is 
supplied, it cannot affect the CNO cycle, i.e. it will not enhance the 
strength of the outburst. The soft X-ray spectrum of 1E~0035.4-7230 ressembles
that of the symbiotic nova RX~J0048.4-7332 in the SMC (Jordan et al. 1996). 
For this source a strong C~{\sc v} edge is also required in the spectral 
modelling. We applied to the {\sl ROSAT} {\sl PSPC} spectrum of this source a 
blackbody spectrum including a C~{\sc v} and a C~{\sc vi} edge and find 
${\rm \tau_{\rm C~V}\ge 5.8}$ and ${\rm \tau_{\rm C~VI}\ge 0.57}$ (68\% 
confidence). The best-fit blackbody temperature is $\sim$$4\ 10^5$~K, somewhat
above the temperature derived from the non-LTE fit by Jordan et al. (1996). 
These authors find a strong stellar wind with a mass-loss rate of $\sim$${\rm 
10^{-6}\ M_{\odot}\ yr^{-1}}$ for the symbiotic nova. 
  
\subsection{Constraints inferred from the Saha equation}

Using the Saha equation we can predict the ratio of the depth in the
C~{\sc v} and C~{\sc vi} edges assuming realistic values for the 
electron density in 
hot atmospheres of WDs and compare this ratio with that
derived from observation. Considering only the ground state of 
ions, the Saha equation is

\begin{equation}{\rm
  \frac{N^+}{N} N_e = \big(\frac{2\pi m_e kT}{h^2}\big)^{3/2}\ 
  \frac{2 g^+}{g}\ e^{-\chi_i/kT}.} 
\end{equation}

Here N is the particle density and $+$ designates the ionized state, 
${\rm N_e}$ is the electron density, g and ${\rm g^+}$ are the statistical 
weights of the ions, and ${\rm \chi_i}$ is the ionization energy. For 
C~{\sc v} and C~{\sc vi} these weights are ${\rm g_{C~{\sc V}}=1}$ and 
${\rm g_{C~{\sc VI}}=2}$ with ${\rm \chi_{C~{\sc V}-C~{\sc VI}} = 6.3\ 
10^{-17}\ J = 392~eV}$. We assume for the optical depth $d\tau = N \sigma ds$ 
and $d\tau^+ = N^+\sigma^+ds^+$. $\sigma$ is the threshold photoionization 
cross-section and ds is the radial distance travelled by a photon. The 
definition of an absorption edge in Section 3.1 allows us to put $ds = ds^+$ 
since all blackbody radiation passes through the same absorbing slab of 
material. Then we get

${\rm d\tau = N\sigma ds}$ 
and ${\rm d\tau^+ = N^+\sigma^+ ds}$ gives

\begin{equation}{\rm
  \frac{N^+}{N} \approx \frac{\tau^+\sigma}{\tau \sigma^+}}
\end{equation}

and 

\begin{equation}{\rm
  N_e = \frac{\tau\sigma^+}{\tau^+\sigma}\big(\frac{2\pi m_e kT}{h^2}
  \big)^{3/2} \frac{2g^+}{g} e^{-\chi_i/kT}}
\end{equation}

with $\rm \sigma=4.7\ 10^{-19}\ cm^{-2}$ and $\rm \sigma^+=1.810^{-19}\ 
cm^{-2}$ (Verner et al. 1996). 

For C~{\sc v} and C~{\sc vi} the electron density can be written as  

\begin{equation}{\rm
  N_e = 1.1710^{23}\ T_5^{3/2} \frac{\tau}{\tau^+} 
  e^{-\chi_i/kT}\ cm^{-3}}
\end{equation}

with the temperature ${\rm T_5}$ in units of $10^5$~K. 
Constraining the temperature
to be in the range $(4-6)\,10^5$~K gives

\begin{equation}{\rm
  N_e = (0.11-8.8)\,10^{20}\ \frac{\tau}{\tau^+}\ cm^{-3}.}
\end{equation}

For high gravity (log~g=9.0) WD atmospheres electron densities in the range 
(${\rm 10^{19}-10^{20}\ cm^{-3}}$) are expected. Such densities are achieved 
for ratios in the range ${\rm \frac{\tau}{\tau^+}}$ ${\rm \sim0.1-1}$. The 
range derived from the blackbody and edge fit ${\rm \frac{\tau}{\tau^+}}$
${\rm \sim0.1-2}$ is therefore consistent with that predicted.
 
\subsection{The evolutionary state of 1E~0035.4-7230}

The temperature and luminosity derived from the different spectral models
can be compared with the asymptotic temperature of $3.25\ 10^5$~K and the 
bolometric luminosity of $\approxgt$${\rm 10^{37}\ erg\ s^{-1}}$ predicted 
by Sion \& Starrfield (1994) for a low-mass (${\rm M_{\rm WD}\sim 0.6-0.7\ 
M_{\odot}}$) WD accreting at a rate of $\sim$${\rm 10^{-8}\ M_{\odot}\ 
yr^{-1}}$ and burning hydrogen steadily for thousands of years. With the 
non-LTE model a kT of $(3.25 \pm 0.16)\ 10^5$~K and a bolometric luminosity 
of ${\rm (1.0-1.2)\ 10^{37}\ erg\ s^{-1}}$ are derived, in agreement with the 
predictions from the Sion \& Starrfield model. The LTE model gives a higher 
effective kT of $5.5\pm0.2\ 10^5$~K and a somewhat lower bolometric luminosity
of ${\rm (3-7)\ 10^{36}\ erg\ s^{-1}}$. Such a temperature is consistent with 
a more massive WD ${\rm M_{\rm WD}\sim 1.1\ M_{\odot}}$, assuming the 
stability relation in Iben (1982) (see Kahabka 1998). However the predicted 
luminosity is at least a factor of 10 too low in order to be explained by 
steady nuclear burning. This could be either due to obscuration of the source 
which is seen at a large inclination of $\sim$40\degree\ (cf. van Teeseling 
et al. 1998), or due to reduced nuclear burning. The X-ray luminosity of the 
source appeared not to have varied by more than $\sim1.30 \pm 0.3$ between 
the ROSAT and BeppoSAX observations. We cannot exclude that this change in 
luminosity could be of instrumental nature due to the uncertain 
cross-calibration of the BeppoSAX LECS and the ROSAT PSPC instruments. It 
also may be due to changing X-ray scattering in the binary system. If it is 
due to a long-term change in the luminosity caused by a cooling of the WD 
envelope then a temperature change by a factor of $1.07\pm 0.05$ would be 
required to account for the derived change in luminosity. 

The extreme softness of the spectrum of 1E~0035.4-7230 allows constraints to 
be set on the existence of a spectrally hard component in this system. Such a 
component may be expected if there is a strong colliding wind present. A wind 
with a mass-loss rate of $\sim$${\rm 10^{-6}\ M_{\odot}\ yr^{-1}}$ with a 
terminal velocity of $\sim$${\rm 500\ km\ s^{-1}}$ from a low-mass WD and a 
mass-loss rate of $\sim$${\rm 10^{-7}\ M_{\odot}\ yr^{-1}}$ with a lower 
terminal velocity from the donor star (as e.g. proposed by van Teeseling \& 
King 1998) should give rise to a colliding wind. We do not detect such a 
component from the BeppoSAX observation. We derive a 90\% confidence upper 
limit luminosity for a bremsstrahlung component with a temperature of 0.5~keV 
(assuming a distance of 60~kpc) of ${\rm 2\ 10^{35}\ erg\ s^{-1}}$. This upper
limit is too large to compare with the luminosity of colliding winds observed 
e.g. by Jordan et al. (1994) in the galactic symbiotic nova RR~Tel of 
$\sim$${\rm 2\ 10^{32}\ erg\ s^{-1}}$.

\begin{acknowledgements}
This research was supported in part by the Netherlands Organization for
Scientific Research (NWO) through Spinoza Grant 08-0 to E.P.J. van den Heuvel.
The \sax\ satellite is a joint Italian--Dutch programme. We thank the staff
of the \sax\ Science Data Center for their support. 
\end{acknowledgements}


\begin{thebibliography}{}
\bibitem{}
Asai K., Dotani T., Nagase F., et al., 1998, ApJ 503, L143
\bibitem{}
Cowley A.P., Schmidtke P.C., Crampton D., Hutchings J.B., 1998, ApJ 504, 854
\bibitem{}
Crampton D., Hutchings J.B., Cowley A.P., et al., 1997, ApJ 489, 903
\bibitem{}
Hartmann H.W., Heise J., 1997, A\&A 322, 591
\bibitem{}
Hartmann H.W., Heise J., Kahabka P., Motch C., Parmar A.N., 
1999, A\&A (in press)
\bibitem{}
Heise J., van Teeseling A., Kahabka P., 1994, A\&A 288, L45
\bibitem{}
Iben I.Jr., 1982, ApJ 259, 244
\bibitem{}
Jordan S., M\"rset U., Werner K., 1994, A\&A 283, 475
\bibitem{}
Jordan S., Schmutz W., Wolff B., et al., 1996, A\&A 312, 897
\bibitem{}
Kahabka P., Pietsch W., Hasinger G., 1994, A\&A 288, 538 
\bibitem{}
Kahabka P., 1996, A\&A 306, 795
\bibitem{}
Kahabka P., Ergma E., 1997, A\&A 318, 108
\bibitem{}
Kahabka P., 1998, A\&A 332, 189
\bibitem{}
King A.R., van Teeseling A., 1998, A\&A 338, 965
\bibitem{} 
Orio M., Della Valle M., Massone G., 1994, A\&A 289, L11
\bibitem{} 
Pagel B.E.J., 1993, Stellar vs. Interstellar Abundances in the Magellanic 
Clouds, in: Lecture Notes in Physics 416, New Aspects of Magellanic Cloud 
Research, eds. Baschek B., Klare G., Lequeux J., 330
\bibitem{}
Parmar A.N., Martin D.D.E., Bavdaz M., et al., 1997a, A\&AS 122, 309
\bibitem{}
Parmar A.N., Kahabka P., Hartmann H.W., et al., 1997b, A\&A 323, L33
\bibitem{}
Parmar A.N., Kahabka P., Hartmann H.W., et al., 1998, A\&A 332, 199
\bibitem{}
Parmar A.N., Oosterbroek T., Orr A., et al., 1999, A\&AS (submitted)
\bibitem{}
Seward F.D., Mitchell M., 1981, ApJ 243, 736
\bibitem{}
Sion E.M., Starrfield S.G., 1994, ApJ 421, 261
\bibitem{} Stanimirovic S., Staveley-Smith L., Dickey J.M., et al., 1999,
           MNRAS 302, 417
\bibitem{}
Van den Heuvel E.P.J., Bhattacharya D., Nomoto K., Rappaport S.A., 1992,
A\&A 262, 97
\bibitem{}
Van Teeseling A., Reinsch K., Pakull M.W., et al., 1998, A\&A 338, 947
\bibitem{}
Van Teeseling A., King A.R., 1998, A\&A 338, 957
\bibitem{}
Verner,~D.A., Ferland,~G.J., Korista,~K.T. \& Yakovlev,~D.G. 1996, ApJ 465, 
487
\end{thebibliography}
\end{document}